\documentclass{aa}
\usepackage{epsfig}
\usepackage{graphicx}
\usepackage{longtable}
\usepackage{longtable} 
\begin{document} 

%\thesaurus{23(03.13.2; 03.19.3; 04.01.1; 13.21.2)}

\title{On the Wilson--Bappu relationship in the \ion{Mg}{ii} k line}

\author{
Angelo Cassatella\inst{1,2}
\and  Aldo Altamore\inst{2}
\and Massimo Badiali\inst{1}
\and Daniela Cardini\inst{1}
} 

%\offprints{Daniela Cardini, 
%\email{cardini at ias.rm.cnr.it}}

\institute
{
Istituto di Astrofisica Spaziale, CNR, Area di Ricerca Tor
Vergata, Via del Fosso del Cavaliere 100, 00133 Roma, Italy 
\and
Dipartimento di Fisica E. Amaldi, Universit\`a degli Studi Roma
Tre, Via della Vasca Navale 84, 00146 Roma, Italy
}

\date{Received / Accepted}

\authorrunning{A. Cassatella et al.}

\titlerunning{The \ion{Mg}{ii} Wilson--Bappu relationship}

\abstract{An investigation is carried out on the Wilson--Bappu effect
in the \ion{Mg}{ii} k line at 2796.34 \AA. The work is based on a
selection of 230 stars observed by both the {\it IUE} and {\it
HIPPARCOS} satellites, covering a wide range of spectral types (F to
M) and absolute visual magnitudes (-5.4$\le$ $M_{V}$ $\le$9.0). A
semi--automatic procedure is used to measure the line widths, which
applies also in presence of strong central absorption reversal. The
Wilson--Bappu relationship here provided is considered to represent an
improvement over previous recent results for the considerably larger
data sample used as well as for a proper consideration of the
measurement errors. No evidence has been found for a possible
dependence of the WB effect on stellar metallicity and effective
temperature.  
\keywords{Stars: distances -- Stars: late-type --
Ultraviolet: general -- Line: profiles -- Catalogs} }
\maketitle
\section{Introduction}
\label{sec:intro}
Several papers have been devoted to the study of the correlation
between the visual absolute magnitude $M_{V}$ of intermediate and
late--type stars, and the width of the \ion{Mg}{ii} k 2796.34  \AA\
emission line (cf. \cite{garcia}; \cite {vladilo}; \cite{elgaroy1}; 
\cite{elgaroy2}; \cite{scoville}, hereafter SMW; \cite{elgaroy3}, 
hereafter EEL).

The width--luminosity relationship for the \ion{Mg}{ii} k line is
qualitatively similar to the one found by Wilson \& Bappu (1957) for
the \ion{Ca}{ii} k optical line (see also recent work by
\cite{wallerstein}), currently known as the Wilson--Bappu effect
(hereafter WB).  Given the similar excitation conditions of these
lines, the dependence of their line widths on stellar luminosity or on
other fundamental stellar parameters rests on the same physical
processes. For a short review on models see Montes et al.  (1994),
references therein, and the recent work by \"Ozeren et al. (1999).

In this paper we study the WB correlation for the \ion{Mg}{ii} k line
using the widest possible sample of stars observed at high resolution
with {\it IUE}, and having a reliable parallax determination from the
{\it HIPPARCOS} experiment.
\noindent More precisely, the purposes of the present paper are:

\noindent
a) to refine the previous luminosity calibrations of the \ion{Mg}{ii}
k line width by using an homogeneous set of measurements of a statistically
significant number of stars covering also spectral types and
luminosity classes poorly represented in previous investigations. A
particular care has been taken to properly consider the measurement
errors.

\noindent
b) to clarify the problem of the possible dependence of the \ion{Mg}{ii} k
line width on stellar effective temperature and metallicity discussed 
in previous papers.

Details about the definition of the present data sample, the
observations, and the method used to measure the line widths are given
in Sections \ref{sec:sample} and \ref{sec:observa}. The results and
the conclusions are presented in Sections \ref{sec:results} and
\ref{sec:conclusi}, respectively.

\section{The present sample of stars}
\label{sec:sample}

A search for all stars with spectral types F, G, K and M with a known
parallax from the {\it HIPPARCOS} experiment and observed at high
resolution in the long wavelength camera by {\it IUE} provides 835
{\it IUE} spectra of 376 stars in a wide range of luminosity from main
sequence to supergiants.  From this initial sample we have purposely
excluded some special classes of stars.  Chromospherically active
binary stars such as RS CVn and BY Dra stars were excluded since their
\ion{Mg}{ii} profiles may be systematically broader than in
non--active stars (\cite{montes}; \cite{ozeren}). Mira variables and
Cepheids were excluded because their V magnitudes and \ion{Mg}{ii}
line profiles are variable and phase--dependent, and also because of
the importance of shock waves associated to pulsation as a source of
line excitation in these objects (see \cite{gillet}; \cite{kraft}).
Also excluded were binary stars having the \ion{Mg}{ii} lines
broadened by peculiar processes (see for example the case of the
M--type giant Mira Ceti, in which the emission lines arise from an
optically thin disk around the white dwarf companion, as reported by
Cassatella et al. (1985) and Reimers and Cassatella (1985)). Finally,
we have excluded rapid rotators and very active stars.

After a further selection based on data quality (see next Section), we
obtained a final sample of 303 spectra of 230 stars.  The sample
includes 11 F--type stars, 56 G--type stars, 133 K--type stars and 30
M--type stars. The range in absolute magnitude covered is about -5.4
$\le$ $M_{V}$ $\le$ 9.0. The most abundant are giant stars (about
50\%) and main sequence stars (25\%). The rest of the sample is
represented by Class I,~II and IV stars, about equally distributed.

\section{Observations and data reduction}
\label{sec:observa}

\subsection{The {\it IUE} spectra}
  
The {\it IUE} high resolution long wavelength spectra have been retrieved
from the {\it INES} (IUE Newly Extracted Spectra) system through its
Principal Centre at {\tt http://ines.vilspa.esa.es}.  A full
description of the {\it INES} system for high resolution data is given in
Cassatella et al. (2000) and Gonz\'alez-Riestra et al. (2000). The
spectra were inspected individually in the \ion{Mg}{ii} k  region to
identify and reject noisy and overexposed or underexposed data.

\begin{figure}
\begin{center}
\includegraphics[width=8.8cm]{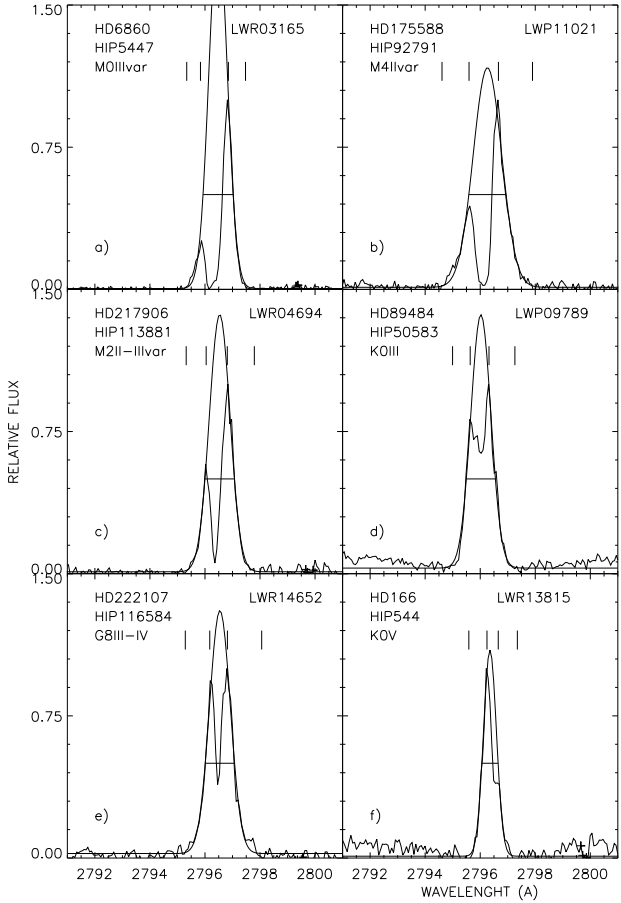}
\caption{Typical profiles of the \ion{Mg}{ii} k line. The vertical marks define
the left and right wings of the emission line. The horizontal line 
indicates the FWHM of the profile}
\label{fig:esempi}
\end{center}   
\end{figure}

Representative examples which show the different
typologies of the \ion{Mg}{ii} k profiles are given in Fig. \ref{fig:esempi}.
One can easily realise that the profiles differ significantly from
case to case. The combined effects of the width of the emission
component, and of the strength and wavelength position of the
absorption cause, quite frequently, the widths of the emission not to
be measurable directly on the line profile itself.

We have then adopted a procedure which allows line widths to be
evaluated in the large majority of cases. It is based on the
reconstruction of the emission profile by fitting the observed
portions of the emission line wings with a gaussian function.  Let the
wavelength boundaries of the violet and red wings to be fitted be
$\lambda_1$ to $\lambda_2$ and $\lambda_3$ to $\lambda_4$,
respectively. These wavelengths were determined manually from
individual spectra and are indicated, in the order, with vertical bars
in the examples of Fig. \ref{fig:esempi}.  The spectra were re--binned
to increase the number of data points by a factor of 5, and slightly
smoothed.  After subtraction of the local continuum and normalization
to the line peak emission, all spectra, irrespective of the typology
of the profile were fitted with a gaussian function.  Line widths were
defined as the full width of the {\it fitted} profile measured at half
maximum of the {\it observed} line peak.  
%%%%%%%%%%
{\bf 
This procedure is
strictly needed for a correct evaluation of the line widths in
presence of strongly mutilated profiles, and is anyhow useful to
improve the accuracy of the line width measurements when dealing with
noisy spectra. In the normal case of 'well behaved' profiles such as c),
d) and e) in Fig. \ref{fig:esempi}, in which a direct measurement of
the line width can be obtained from the spectra, the procedure
provides values wich are fully consistent, within the errors, with the
ones obtained through the traditional methods.
}
%%%%%%%%%%%%%%%%

Examples of extreme, although quite frequent cases in which a direct
measurement using traditional methods cannot be done, are the profiles
a) and b) of Fig. \ref{fig:esempi}, in which the central absorption
divides the profile in two unequal parts, being the flux peak of one of
them less than 50\% of the other, or in profiles of the type shown in
panel f), in which one of the wings is mutilated by superimposed
absorption (on the red wing in this case).

%%%%%%%
{\bf 
It should be stressed that, according to the numerical simulations by
Cheng et al. (1997), the fluxes and profiles of chromospheric
resonance lines depend on the temperature structure and dynamics of
the atmosphere, which may inevitably lead to line profiles that are
rather different from gaussian.  Our choice to use a gaussian profile
is just  an empirical approach, having the only purpose
to improve the accuracy of the line width measurements through
a reliable reconstruction of the emission line wings. This
procedure is particularly useful in presence of severe intrinsic 
or interstellar line absorption.
}

%%%%% REFEREE
{\bf
As a general rule, the results of the fit were rejected whenever the
peak intensity of the fitted profile exceeded the peak of the
\ion{Mg}{ii} k observed profile by more than a given amount, somewhat
arbitrarily fixed to 180\%. 
Also rejected were poor quality results where the
relative error of the fit was larger than 10\% (r.m.s.)  and the
results from strongly mutilated and asymmetric line profiles with a
deep ``central'' absorption dividing the profile in two halves, with
one having a peak intensity less than 10\% of the other (i.e. similar
to the limiting case a) of Fig. \ref{fig:esempi}).  
}
%%%%%%%%%%

As for the errors on measured line widths, we have estimated it to be
approximatively equal to half of the sampling interval of the spectra,
corresponding to 3.64 km~s$^{-1}$, for line profiles as in
Fig. \ref{fig:esempi} c), d) and e), and twice this value for the
other typologies. 

The measured line widths $W$ need to be corrected for instrumental
broadening.  Assuming the {\it IUE} Point Spread Function for high
resolution spectra to be a gaussian with a full width at half maximum
b=18 km~s$^{-1}$ at 2800 \AA\ (\cite{evans}), the true width
$W_0$ can be written as:
  
\begin{equation}
\rm{W_{0}^2 = {W}^2 -b^2}
\label{eq:corr}
\end{equation}
This correction, which is especially important for narrow lines, has
apparently not been taken into account in some previous works on the
subject (see Section 4.4). Whenever multiple spectra were available
for the same star, the line width was computed as the weighted mean of
the individual measurements.

\subsection{The Hipparcos data}
For each of the 230 stars, the following information has been
extracted from the Hipparcos Catalogue (\cite{esa}): the Johnson V
magnitude, the $B-V$ colour and the corresponding error, the parallax
with its error, the spectral type and luminosity class, and
information on variability or duplicity.  In a few cases (suitably
labeled in Table A, described below) the $B-V$ colour index, its
error, and the spectral type were taken from the Strasbourgh Data
Center because the values contained in the {\it HIPPARCOS} catalogue
were not considered reliable enough.

Table A\footnote{Table A is only available in electronic form at CDS
via anonymous ftp to cdsarc.u-strasbg.fr (130.79.128.5) or via
http://cdsweb.u-strasbg.fr/Abstract.html} contains, for each object of
the final sample, the HIP and HD numbers, the spectral type and
luminosity class, the absolute magnitude $M_V$ and its error
$\sigma(M_V)$, the $B-V$ colour index with the corresponding error
$\sigma(B-V)$, the {\it IUE} images used, the logarithm of the
\ion{Mg}{ii} k line width $log~W_0$ with its error
$\sigma(log~W_0)$. The absolute magnitudes were computed without any
allowance for interstellar reddening.  The errors in $M_V$ have been
derived from the errors in trigonometric parallaxes assuming a typical
error of $\pm$ 0.01 mag in the observed visual magnitudes.

\begin{figure}
\hspace{0.8cm}
\includegraphics[width=8.8cm]{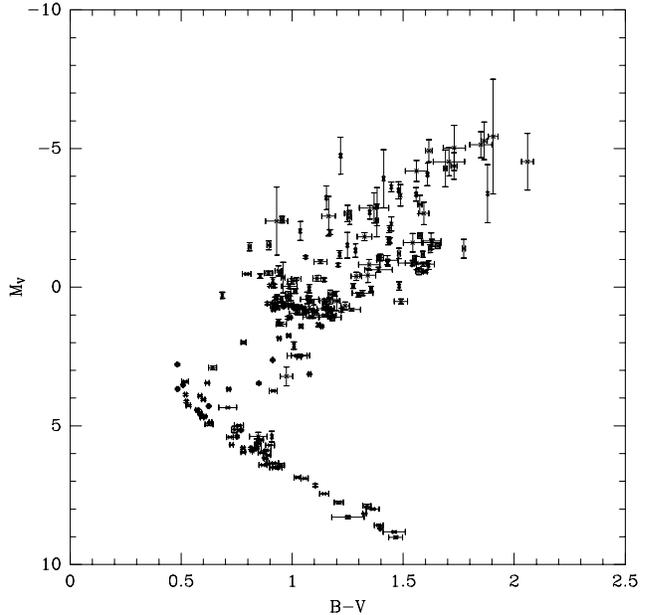}
\caption{Colour--magnitude diagram for the 230 stars in our sample}
\label{fig:hrdiag}
\end{figure}

The $M_V$ vs $B-V$ diagram of our sample of stars is shown in
Fig. \ref{fig:hrdiag}. One can easily recognize that main sequence
stars with $B-V$ $\geq$ 0.5, as well as giant and supergiant stars are
well represented in our sample.

\section{Results}
\label{sec:results}

\subsection{The present \ion{Mg}{ii} k width-luminosity relationship}
\label{sec:present}
To  determine the coefficients $a$ and $b$ of the \ion{Mg}{ii} k Wilson--Bappu
relationship 
 
\begin{equation}
\it {M_{V} = a + b~log W_0}
\label{eq:equation}
\end{equation}

\noindent we applied linear regression algorithms to the $M_V$ and
$log~W_0$ data of the 230 stars in Table A. The numerical codes
available from Press et al. (1992) were used.

We considered regression solutions which take into account the
measurement errors in only one of the variables, as often done in the
literature, or in both variables. 
%The weights were computed as the
%inverse squared standard deviation. 
Unweighted regression lines were also computed, although only for
comparison purposes.  In any case, the correlation coefficient between
the $M_V$ and $log~W_0$ data is $r$=-0.94.

The results obtained in the different cases are reported in Table
\ref{tab:coeff1}.  Column 1 contains the label assigned to the fit. In
Column 2 we specify the variable taken as independent.  The variables
for which the errors were taken into account are indicated in Column
3.  In particular, fits $A$ and $B$ were done without taking into
account the measurement errors; in fits $C$ and $D$ only the errors on
$M_V$ or $log~W_0$ were considered, respectively; in the
fits $E$ and $F$ the errors on both variables were
taken into account.  The regression coefficients $a$ and $b$ together
with the corresponding errors are given in Column 4 and 5.  They are
consistently given in the format appearing in Eq. \ref{eq:equation},
also in the case a switch of variables has been performed. In that
case the uncertainties on the coefficients have been accordingly
propagated.

As shown in Table \ref{tab:coeff1}, the $a$ and $b$ coefficients
derived by switching the independent variable ($log~W_0$ or $M_V$),
vary significantly if allowance is made for the measurement errors
in one variable only.  This is the case of fit $C$, in which the
independent variable is $log~W_0$ and only the errors on $M_V$ are
taken into account, and fit $D$, in which the independent variable is
$M_V$ and only the errors in $log~W_0$ are taken into account.  On the
contrary, fits $E$ and $F$, in which the errors on both variables are
considered, provide fully consistent coefficients, irrespective of the
variable considered as independent, as expected (\cite{akritas}).

The results in Table \ref{tab:coeff1} show the important role played
by the method used to analyse the data. This point is further
discussed in Section \ref{sec:comparison}.

We consider fit $E$ or, equivalently, fit $F$ in Table
\ref{tab:coeff1} as the most representative of the WB relationship,
since the measurement errors on both variables were taken into
account. In conclusion, the proposed WB relationship for the
\ion{Mg}{ii} k line is:

\begin{equation}
\it {M_{V} = (34.56 {\pm} 0.29)- (16.75 {\pm} 0.14)~ log W_0}
\label{eq:bappueq}
\end{equation}

\begin{figure}
\hspace{0.8cm}
\includegraphics[width=8.8cm]{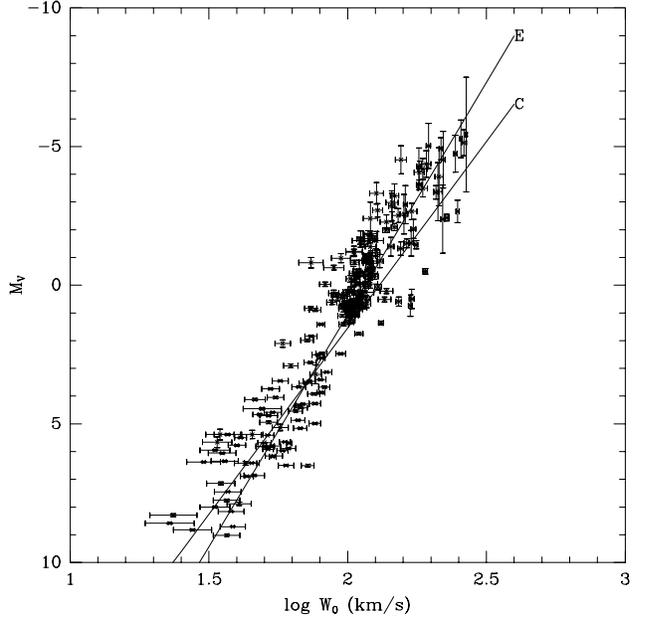}
\caption{The Wilson--Bappu effect in the \ion{Mg}{ii} k line for the 230 stars
in our sample. The regression line labeled E corresponds to the
proposed relationship in Eq. \ref{eq:bappueq}, obtained by taking into
account the measurement errors on both variables. For comparison, also
the regression line C (see Table \ref{tab:coeff1}) is provided, which
has been obtained from the same data by taking into account
only the errors on $M_V$}
\label{fig:bappufig}
\end{figure}

The observed data, together with the corresponding error bars and the
adopted WB relationship (labeled $E$) are shown in
Fig. \ref{fig:bappufig}. For comparison, also fit $C$ is shown
(labeled $C$). 

\begin{table}
\caption{Coefficients of the Wilson-- Bappu relationship}
\begin{flushleft}
\begin{tabular}{l c c c c}
\hline
\hline 
\noalign{\smallskip} FIT&I.V.&$\sigma$ & $a$ & $b$ \\
\noalign{\smallskip}
\hline
\noalign{\smallskip}
A&$log~W_0$   & none      &29.54 $\pm$0.71 &-14.37 $\pm$0.36 \\
B&$M_V$       & none      &33.58 $\pm$0.84 &-16.42 $\pm$0.41 \\
C&$log~W_0$   & $M_V$     &28.42 $\pm$0.02 &-13.44 $\pm$0.01 \\
D&$M_V$       & $log~W_0$ &34.71 $\pm$0.18 &-16.69 $\pm$0.09 \\
E&$log~W_0$   & both      &34.56 $\pm$0.29 &-16.75 $\pm$0.14 \\
F&$M_V$       & both      &34.59 $\pm$0.29 &-16.76 $\pm$0.14 \\
\noalign{\smallskip}		     
\hline				     
\noalign{\smallskip}
\end{tabular}
\par
\vspace{0.3cm}
\noindent
{\it Notes}: Column 2, labeled I.V., specifies the variable, $log~W_0$
or $M_V$, taken as ``independent variable''. Column 3, labeled $\sigma$,
indicates for which of the variables the measurement errors were taken
into account. The coefficients of the linear fit $M_{V} = a + b~
log~W_0$, together with the corresponding errors are given in Columns 4
and 5. The correlation coefficient for all fits is $r$= -0.94.
\label{tab:coeff1}
\end{flushleft}
\end{table}

\subsection{Dependence on Effective Temperature}

We investigated the possible dependence of $log~W_0$ on stellar
effective temperature, an item which has been addressed in previous
recent analyses and is still under discussion.

\begin{figure}
\begin{center}
\includegraphics[width=8.8cm]{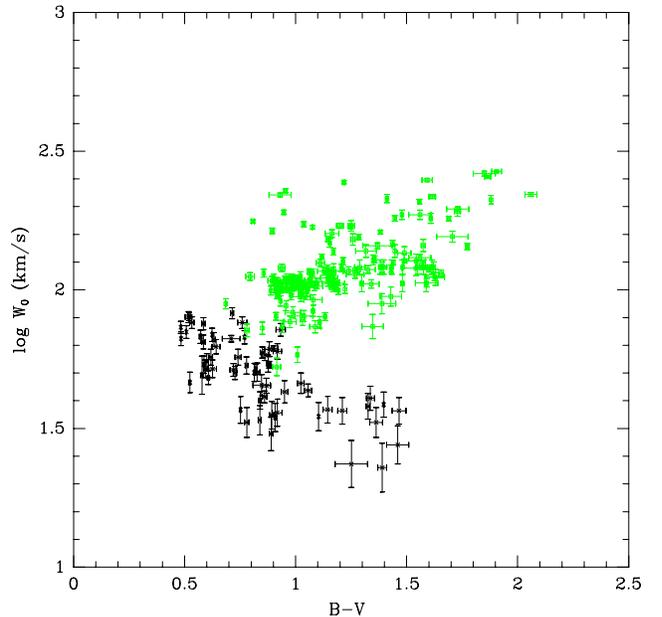}
\caption{Dependence of line width $log~W_0$ on $B-V$ for the stars in
our sample. Black symbols refer to main--sequence stars and grey symbols to
giant and supergiant stars}
\label{fig:w3}
\end{center}   
\end{figure}

Fig. \ref{fig:w3} shows our measurements of $log~W_0$ plotted as a
function of the $B-V$ colour index, taken to be a temperature
indicator. This representation is obviously equivalent to the
colour--magnitude diagram in Fig. \ref{fig:hrdiag} by virtue of the
Wilson--Bappu linear relationship. The figure shows, in particular,
that $log~W_0$ is a decreasing function of $B-V$ for main sequence
stars (indicated as black points), while it is an increasing function
of $B-V$ for giants and supergiants (gray points on the upper part of
the diagram). Therefore, the weak inverse relationship between line
widths and stellar temperature found by SMW (see their Fig. 5) simply
reflects the distribution of their sample of stars in the HR diagram,
which are mostly giants and supergiants.

\begin{figure}
%\begin{center}
\hspace{0.8cm} 
\includegraphics[width=8.8cm]{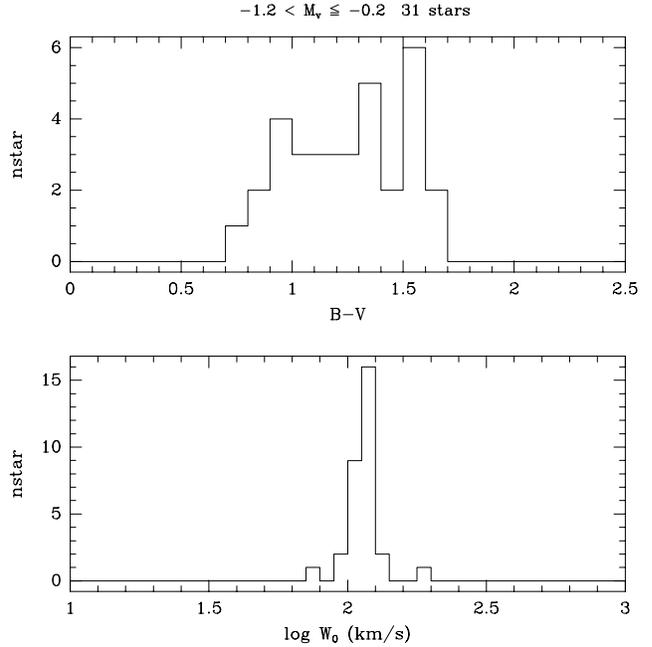}
\caption{Histogram of $B-V$ (top)  and $log~W_0$ for stars in a restricted
range of absolute magnitude -1.2 $<$ $M_{V}$ $\le$ -0.2 (31 stars)}
\label{fig:w6}
%\end{center}   
\end{figure}

An effective way to face the problem is to concentrate on stars in
restricted ranges of $M_V$ (so that any dependence on luminosity is
taken out), and to compare their distribution in $B-V$ and in
$log~W_0$. To start, we selected the stars having -1.2 $<$ $M_{V}$
$\le$ -0.2 (31 stars). The results, reported in Fig. \ref{fig:w6},
show clearly that $B-V$ is distributed over a wide range of values
(0.7 $\le$ $B-V$ $\le$ 1.7), while the distribution in $log~W_0$ has a
very prominent peak around $log~W_0$=2.05.  Line widths appear then to
be substantially insensitive to the colour index, i.e. to the stellar
effective temperature, at least in the above range of $M_V$.  Tests
made with stars lying in different ranges of $M_V$ show a pattern
which is  similar to the one in Fig. \ref{fig:w6}, but the results
cannot be considered as equally conclusive due to the paucity of stars
available, and/or to the narrow $B-V$ range covered.

\begin{figure}
\hspace{0.8cm}
\includegraphics[width=8.8cm]{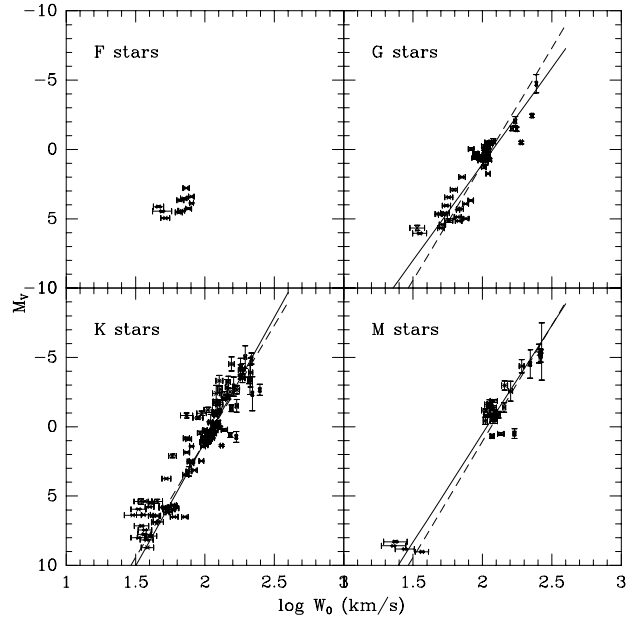}
\caption{
The Wilson--Bappu relationship  for different spectral types}
\label{fig:w4}
\end{figure}

Using a different approach, previously followed by EEL, we
investigated on whether a different width-luminosity relationship
applies to different spectral types. In Fig. \ref{fig:w4}, we compare
our results of the width--magnitude relationships for G, K and M stars
(solid lines) with the regression line corresponding to the full set
of stars in our sample (Eq. \ref{eq:bappueq}; dashed line). The
corresponding coefficients together with the number of stars used are
given in Table \ref{tab:coeff2}.  The fits were done by taking into
account the errors on both variables. No attempt has been made to fit
F--type star's data, which are too few for a statistical approach.  In
any case, not even a better statistics could possibly help in view of
the intrinsically small range of $M_V$ covered by F--type stars.

As it appears from Table \ref{tab:coeff2}, the coefficients for
K--type stars (133 objects), are very similar to the ones in
Eq. \ref{eq:bappueq}, as expected,  given that  these stars
represent more than 50\% of the entire sample.

The poorest statistics is that of M--type stars (30 objects), which
are also unevenly distributed in $M_V$ (see the gap 1
$\le$ $M_{V}$ $\le$ 8). Given these limitations, the WB coefficients
are strongly dependent on the stars considered in input, so that the
differences appearing from the Table are not significant enough.

As for G--type stars (56 objects), we find a significantly flatter
slope than for K-type stars (133 objects).  In particular, the line
widths of the G-type stars in Fig. \ref{fig:w4} with $M_{V}$ $\le$ -1
are broader than one would expect from the WB relation in
Eq. \ref{eq:bappueq}.  A similar trend has been reported by EEL for a
sample of 22 G--type stars.  In this regard, it is interesting to
mention that also Wallerstein et al. (1999), in their study of the WB
effect in the \ion{Ca}{ii} k line, found a group of luminous G--type
stars having substantially broader \ion{Ca}{ii} lines than predicted
by the WB relation. Whether the apparently different WB relation for
G--type stars should be regarded as a true temperature effect as
claimed by EEL, or is due to other causes (including selection
effects), it cannot be clarified at this stage.

\begin{table}
\caption{Coefficients of the fit for different spectral types}
\begin{flushleft}
\begin{tabular}{l c c c c}
\hline
\hline 
\noalign{\smallskip}Sp. Type&N& $a$ & $b$ & $r$\\
\noalign{\smallskip}
\hline
\noalign{\smallskip}
G--stars&56 &28.82 $\pm$0.42 &-13.88 $\pm$0.20 &-0.91 \\
K--stars&133&36.94 $\pm$0.41 &-17.93 $\pm$0.20 &-0.93 \\
M--stars&30 &31.77 $\pm$1.22 &-15.63 $\pm$0.58 &-0.96\\
\noalign{\smallskip}		     
\hline				     
\noalign{\smallskip}
\end{tabular}
\par
\vspace{0.3cm}
\noindent
{\it Notes}: Columns 1 to 5 give, in the order, the spectral type
considered, the number of stars used, the regression coefficients with
the corresponding errors, and the linear correlation coefficient.
\label{tab:coeff2}
\end{flushleft}
\end{table}

\begin{figure}
%\hspace{0.8cm}
%\psfig{file=w5.eps,width=8.8cm}
%\psfig{file=wmetmsgig.eps,width=8.8cm}
\includegraphics[width=8.8cm]{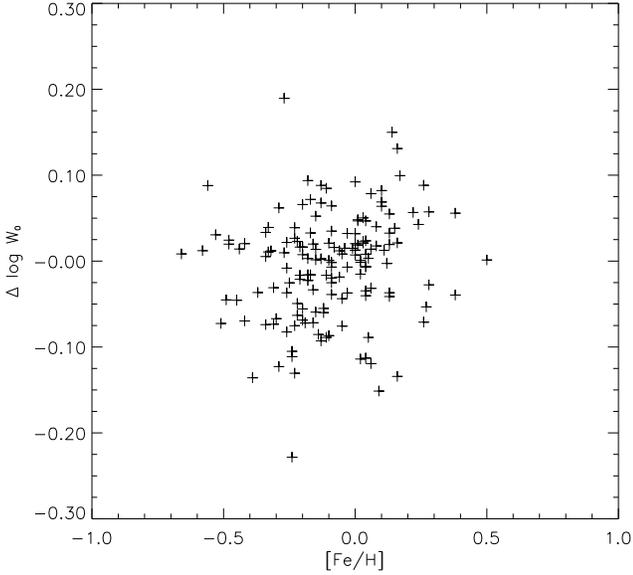}
\caption{The departure of the observed line width $log~W_0$ with
respect to the Wilson--Bappu relationship in Eq. \ref{eq:bappueq} is
plotted as a function of  metallicity $[Fe/H]$ for 156 stars in our sample}
\label{fig:metal}
\end{figure}

\subsection{Dependence on Metallicity}

We investigated on possible departures from the Wilson--Bappu
relationship in Eq. \ref{eq:bappueq} due to a dependence of the
\ion{Mg}{ii} k line width on metallicity. To this purpose, we looked
for $[Fe/H]$ literature data and found information for 156 stars of
our sample.  For 12 stars, the $[Fe/H]$ values are from Carney et
al. (1994), and for the remaining stars from Cayrel de Strobel et
al. (1997). Being this latter a compilation from several sources, more
then one determination was in general available for a given star. In
these cases, the most recent one has been adopted, assumed to be the
most reliable.  In Fig. \ref{fig:metal} we plot as a function of
$[Fe/H]$ the difference $\Delta$$log~W_0$ between the measured line
width $log~W_0$ and the value expected from Wilson--Bappu relationship
in Eq. \ref{eq:bappueq}.  It clearly appears from the figure that, in
spite of the quite large range in metallicity, there is no 
trend for the line widths to show any systematic deviation. The
correlation coefficient between $[Fe/H]$ and the deviations
$\Delta$$log~W_0$ from Eq. \ref{eq:bappueq} is indeed very poor:
$r$=0.16.

As a further confirmation of this point, we divided the 156 stars in
four groups according to their metallicity. For each group we made a
linear fit of $M_V$ as a function of $log~W_0$ (allowing for the
errors on both variables). The metallicity groups were defined as
follows: $log[Fe/H]$ $\le$~-0.3 (23 stars), -0.3 $<$ $log[Fe/H]$
$\le$~-0.1 (56 stars), -0.1 $<$ $log[Fe/H]$ $\le$ +0.1 (54 stars), and
$log[Fe/H]$ $>$ +0.1 (23 stars). Slightly different slopes and
intercepts of the WB relationship were found, but no dependence on
metallicity was  found.

\subsection{Comparison with previous determinations}
\label{sec:comparison}

The results of the present study are compared with the literature
values of EEL and SMW in Table \ref{tab:literature} which gives, for
each author, the number of stars used, the coefficients $a$ and $b$
(with their errors, if available), and the range of $M_V$ values from
which the WB relationship has been derived.  Reference is made only to
papers making use of the same definition of line width (FWHM in this
case), and based on precise {\it HIPPARCOS} parallaxes.

\begin{table}
\caption{Comparison with previous results}
\begin{flushleft}
\begin{tabular}{@{}l c c c l@{}}
\hline
\hline 
\noalign{\smallskip}Reference&N& $a$ & $b$ &$M_V$ range\\
\noalign{\smallskip}
\hline
\noalign{\smallskip}
Eq. \ref{eq:bappueq} (Fit E)&230&34.56$\pm$0.29&-16.75$\pm$0.14&-5.4~~~9.0\\
Fit C               &230&28.42$\pm$0.02&-13.44$\pm$0.01&-5.4~~~9.0\\
EEL   1999         &~65&35.25$\pm$2.17&-17.61$\pm$1.10&-5.1~~12.3\\
%EEL   1999/2         &~65&31.19~~~~~~~~~&-14.66~~~~~~~~~&-5.1~~12.3\\
SMW   1998           &~92&28.63~~~~~~~~~&-13.53~~~~~~~~~&-2.2~~~7.5\\
%Vlad. 1987          &~41&41.36$\pm$0.78&-19.58$\pm$0.40&-1~~~~~7~\\
\noalign{\smallskip}		     
\hline				     
\noalign{\smallskip}
\end{tabular}
\par
\vspace{0.3cm}
\noindent
{\it Notes}: EEL and SMW  stand for Elgar{\o}y et al. (1999)
and Scoville \& Mena--Werth (1998),
% and Vladilo et al. (1987),
respectively. Column 2 provides the number of stars used by the
different authors. Fits $E$ (the proposed WB relationship) and $C$ are the
same as in Table \ref{tab:coeff1}. 
\label{tab:literature}
\end{flushleft}
\end{table}
%Weiler et al.   &49 &34.93~~~~~~~~~&-15.15~~~~~~~~~~&-0.91 \\
%Garc\'ia A. et al.&15+&34.86~~~~~~~~~&-15.14~~~~~~~~~~&-0.95 \\
%Fit C (Tab. \ref{tab:coeff1})&230&28.42$\pm$0.02&-13.44$\pm$0.01&-5.4~~~9.0\\

Since our sample contains 84 stars in common with SMW and 34 in common
with EEL, it is useful to make a direct comparison between the line width
measurements.  

\begin{figure}
\includegraphics[width=8.8cm]{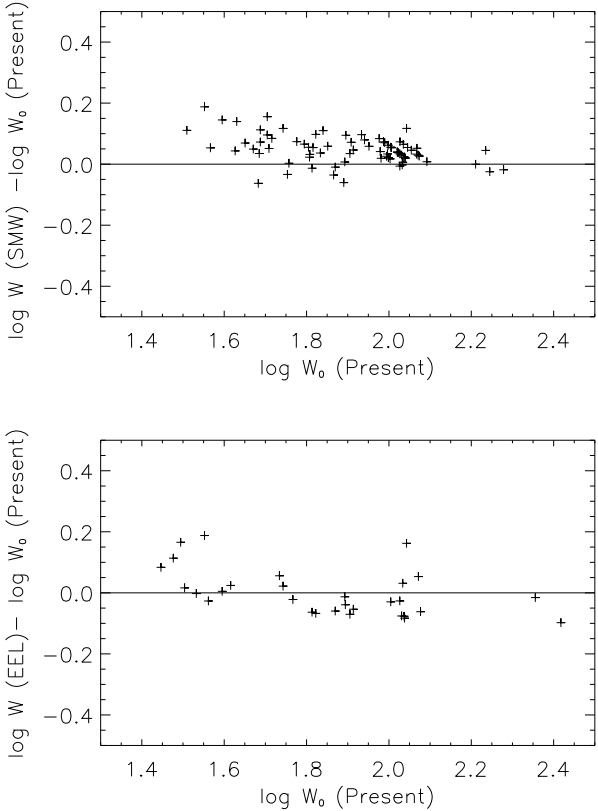}
\caption{Comparison of our line width measurements with those of
Scoville \& Mena--Werth (1998; labeled SMW) and of Elgar{\o}y et al.
(1999; labeled EEL)}
\label{fig:scoville}
\end{figure}

As it appears from the top panel of Fig. \ref{fig:scoville} our
measurements of $W_0$ are in a good agreement with SMW's values for
the broadest lines, but there is an increasing discrepancy for
narrower and narrower lines, whose widths are systematically
overestimated by SMW, as it would happen if their measurements were
not corrected for the instrumental width (Eq. \ref{eq:corr}), an item
which is not specified in their paper.  Compared with ours, the SMW's
measurements of $W$ are about 8\% higher, on average. In spite of this
discrepancy, one can see from Table \ref{tab:literature} that SMW's fit
is very close to our fit $C$, but differs strongly from the
others. This might suggest that it has been obtained by weighting only
on $M_V$, as fit $C$.

The agreement with EEL's measurements of $W$ (Fig.
\ref{fig:scoville}, bottom panel) is substantially better, amounting
to 2\%, if 4 discrepant data points are excluded (note that for such a
comparison, we have averaged the two determinations of $W$ obtained by
EEL with different methods -- see their Table 1).  As one can easily
appreciate from Table \ref{tab:literature}, our best estimate of the
WB effect in Eq. \ref{eq:bappueq} is compatible with EEL's fit, at
least within their large error bars. Unfortunately, no information
about the treatment of measurement errors is given by EEL. The only
authors who declared to have taken the measurement errors on both
variables into account are Vladilo et al. (1987), but their results
cannot be compared with ours due to their much more limited range in
$M_V$ (-1 to 7).  We should note also that the values of $M_V$ here
adopted, coincide with those of SMW, while there is, on average, a
systematic difference of 0.12 mag, of unknown origin, between EEL's
values of $M_V$ and ours.

In conclusion, given the substantially larger amount of observations
analysed in this work, and the proper way to take the measurement
errors into account, we consider Eq. \ref{eq:bappueq} to adequately
represents the WB effect for the \ion{Mg}{ii} k line, at least within
the observational limits set by the {\it IUE} and {\it HIPPARCOS}
experiments.

\section{Conclusions}
\label{sec:conclusi}

In this paper we provide an accurate Wilson--Bappu relationship for
the \ion{Mg}{ii} k line (see Eq. \ref{eq:bappueq}) based on {\it IUE}
and {\it HIPPARCOS} observations, which can be used to estimate
stellar distances in a wide range of absolute magnitudes: -5.4 $\le$
$M_{V}$ $\le$ 9.0. The sample of stars (230 objects) is considerably
larger than the ones used in previous works. 

It has been shown that the coefficients of the WB relationship
critically depend upon the choice of the method of analysis.  In our
case, different fitting algorithms have been tested and
the one taking into account the measurement errors on both variables
has been adopted as the most suitable. 

In spite of the more accurate method and of the wider sample of stars
used, we have not found any observational confirmation to the claimed
dependence of the WB effect on effective temperature.  On the
contrary, we have found arguments against such an effect, as the one
provided in the histograms in Fig. \ref{fig:w6}, which show that the
\ion{Mg}{ii} line width is substantially the same for stars in a
narrow range of $M_V$ (-1.2 $<$ $M_{V}$ $\le$ -0.2), in spite of the
wide range of $B-V$ values. In any case, on the basis of theoretical
considerations, the temperature effect is expected to be small
(\cite{reimers}, \cite{ozeren}).

We also investigated on possible systematic deviations caused by
differences in metallicity. Using metallicity data for 156 stars in
our sample, we came to the conclusion that such an effect is not
present, at least within the accuracy of the measurements.

Given the difficulty to disetangle the metallicty and temperature
effects using the present sample of field stars, it would be important
to extend the analysis to homogeneous populations of stars such as
galactic clusters.  More in general, observations by newly proposed
space missions (e.g. GAIA) will hopefully enrich our knowledge on
parallaxes of very large numbers of stars and clarify the problems
left open.
  
\begin{acknowledgements}
We would like to thank Prof. M.J. Fern\'andez--Figueroa and Drs. Anna
Marenzi, Vittoria Caloi, and R. Gonz\'alez--Riestra for useful
comments and advice.

\end{acknowledgements}

\end{document}